\documentclass[%
 aip,
 amsmath,amssymb,
 reprint,%
]{revtex4-1}

\usepackage{graphicx}% Include figure files
\usepackage{dcolumn}% Align table columns on decimal point
\usepackage{bm}% bold math
\usepackage[utf8]{inputenc}
\usepackage[T1]{fontenc}
\usepackage{mathptmx}
\usepackage{etoolbox}

\renewcommand{\vec}[1]{\bm{#1}}

\makeatletter
\def\@email#1#2{%
 \endgroup
 \patchcmd{\titleblock@produce}
  {\frontmatter@RRAPformat}
  {\frontmatter@RRAPformat{\produce@RRAP{*#1\href{mailto:#2}{#2}}}\frontmatter@RRAPformat}
  {}{}
}%
\makeatother
\usepackage{}	% required for `\align' (yatex added)
\begin{document}

\preprint{AIP/123-QED}

\title[]{Turbulent scaling law in Ogata K\=orin's \textit{Red and White Plum Blossoms}}
% Force line breaks with \\
\author{Takeshi Matsumoto}
\email{takeshi@kyoryu.scphys.kyoto-u.ac.jp}
\affiliation{ 
Division of Physics and Astronomy, Graduate School of Science, Kyoto University\\%\\This line break forced with \textbackslash\textbackslash
Kitashiarakawa Oiwakecho Sakyoku Kyoto, 606-8502, Japan.
}%

\date{\today}% It is always \today, today,

\begin{abstract}
Stylized turbulent swirls depicted in artworks are often analyzed with the modern 
tools for real turbulent flows such as the power spectrum and the structure function. 
Motivated by the recent study on \textit{The Starry Night} of van Gogh
(Ma \textit{et al}., Phys. Fluids, \textbf{36} 095140, 2024), 
we here analyze Ogata K\=orin's \textit{Red and White Plum Blossoms}, in particular
its swirling pattern and the bark of the plum-tree trunk. 
The results show that they follow closely the Obukhov--Corrsin spectrum $k^{-5/3}$
in the inertial-convective range of the passive scalar advected by the homogeneous and isotropic turbulence. 
Furthermore their 4th- and 6th-order structure functions exhibit approximately the same intermittent
scaling law of the passive scalar. We discuss several possible explanations of this consistency.
\end{abstract}

\maketitle

%\begin{quotation}
%The ``lead paragraph'' is encapsulated with the \LaTeX\ 
%\verb+quotation+ environment and is formatted as a single paragraph before the first section heading. 
%(The \verb+quotation+ environment reverts to its usual meaning after the first sectioning command.) 
%Note that numbered references are allowed in the lead paragraph.
%%
%The lead paragraph will only be found in an article being prepared for the journal \textit{Chaos}.
%\end{quotation}

\section{\label{s:intro}Introduction}
Great artworks depicting fluid flows compel some fluid dynamicists
to analyze them as they do for real flows
in laboratory experiments, field measurements,
and numerical simulations.  Irresistible examples include 
Leonardo da Vinci's drawings\cite{mona14, ma_leo21,colagrossi_da_2021},
\textit{The Great Wave off Kanagawa} by Katsushika Hokusai\cite{cart09,dudley13}, 
and \textit{The Starry Night} by van Gogh\cite{aragon08,beattie19,finlay20,ma24}, 
just to name a few.

Da Vinci's most celebrated drawing of turbulence  \textit{Studies of Water}, which depicts 
water from a duct falling into a larger pool and causing many air bubbles, 
is now numerically reproduced with the smoothed particle hydrodynamics 
method\cite{mona14, colagrossi_da_2021}.
What is remarkable is that the attentions are 
drawn not only to realistic expressions of flows, but also to stylized ones, 
such as the works by Hokusai and van Gogh. 
\textit{The Great Wave off Kanagawa} is compared with 
a photograph of a certain type of oceanic breaking waves taken from a 
research vessel, showing consistency of 
his distinctive expression with the physics\cite{dudley13}.

\textit{The Starry Night} is well-known for the magnificent swirls in the sky at night. 
Those stylized swirls have been analyzed by calculating the power spectrum density
of the luminance of the image \cite{beattie19,finlay20,ma24}. 
It is considered that the luminance
corresponds to the passive scalar advected by the real turbulent flow \cite{aragon08,ma24}.
Indeed early studies find power-law scaling in the luminance spectrum.
However its scaling exponent is a matter of debate\cite{beattie19,finlay20,ma24}. 
A recent detailed
study\cite{ma24} shows that the luminance spectrum of the swirls 
is composed of $k^{-5/3}$ and $k^{-1}$ parts, where $k$ is the wavenumber.  
They argue that the first and second parts correspond to the Obukhov--Corrsin spectrum and
the Batchelor spectrum of the passive scalar turbulence at high Schmidt numbers\cite{ma24}.
The Batchelor-spectrum part is confirmed also by the second-order 
structure function of the luminance\cite{ma24}.
Therefore, the iconic swirls in \textit{The Starry Night} respect the physical law
of turbulent flows.

Motivated by this recent study\cite{ma24}, we here analyze 
stylized turbulent swirls and another part in a Japanese painting in the 18th century 
by utilizing the spectrum and structure function.
The painting is \textit{Red and White Plum Blossoms} by 
Ogata K\=orin (1658--1716), which is shown in Fig.~\ref{f:pic}.
It is less known than
\textit{The Great Wave off Kanagawa} by Hokusai (1760--1849). 
Nonetheless, \textit{Red and White Plum Blossoms}  is considered 
also as one of the most important
Japanese artworks of the Edo period (1603--1868).
The artists influenced by the style of K\=orin and his predecessors 
are now called ``Rin-pa''\cite{moma} coined from his given name.
It is known that Hokusai is one of them (see e.g., Refs.~\onlinecite{yasui96, fujisawa22}).

Although the title \textit{Red and White Plum Blossoms} does not mention it, 
the river is flowing between the red plum tree on the right and the white plum tree on the left.
The flow is expressed by boldly stylized swirls that 
exhibit a irregular hierarchical structure. This intriguing hierarchy leads us to 
identify its scaling law. Furthermore we analyze also the bark of the trunk of the white plum tree.
In particular, we first show that the luminance spectrum follows closely $k^{-5/3}$, which
can be interpreted as the Obukhov--Corrsin spectrum\cite{o49,co51,tl72} of the passive scalar
advected by the homogeneous and isotropic turbulence.
Then we demonstrate that the luminance obeys approximately the same intermittent
scaling law of the passive scalar turbulence\cite{rmp,w00,wg04,gy13} by calculating the $4$th- and $6$th-order 
structure functions. Namely, the power-law exponents of the $p$-th order structure functions
of the luminance in the scaling range are different from $p/3$ ($p=4$ and $6$ here), but approximately 
the same as those of the passive scalar.  
To the best of our knowledge, such consistency of an artwork with the real turbulent 
flow including the intermittency is not known. For the stylized swirls, it may appear
conceivable, but the same consistency holds also for the the depicted bark pattern on the plum 
tree. 

Why are the flow pattern and the bark pattern on the plum tree 
in \textit{Red and White Plum Blossoms} 
created about 300 years ago quantitatively
consistent with the detailed laws of the present-day fluid dynamics?
Is it just a coincidence? Although we do not have an answer to this question yet, 
we will discuss several possible explanations at the end of this paper
together with the insights\cite{dudley13, ma24} obtained from
\textit{The Great Wave off Kanagawa} and \textit{The Starry Night}. 

The organization of the paper is as follows.
In Sec.\ref{s:pre} we explain how we pre-process the electronic image 
of the painting.
In Sec.\ref{s:spc} we analyze the luminance spectrum of the swirls.
In Sec.\ref{s:stf} we calculate the structure function of the luminance.
Concluding discussions are given in Sec.\ref{s:concl}.

\begin{figure*}
\includegraphics[width=0.9\textwidth]{Ogata_Korin_-_RED_AND_WHITE_PLUM_BLOSSOMS_National_Treasure_-_Google_Art_Project.jpg} 
\caption{\label{f:pic} Ogata K\=orin's \textit{Red and White Plum Blossoms}, a collection of the MOA Museum of Art in Atami, Japan.
This is a folding screen consisting of two separate panels. The black column in the center of this image is a margin between the two panels. The dimension of each panel is $156 {\rm cm}$ (height) $\times 172.2 {\rm cm}$ (width). The panels can be folded in the middle.
This particular image (JPG format) provided by Google Art \& Culture 
is available at Wikimedia Commons ({\scriptsize \texttt{https://commons.wikimedia.org/wiki/File:Ogata\_K\=orin\_-\_RED\_AND\_WHITE\_PLUM\_BLOSSOMS\_(National\_Treasure)\_-\_Google\_Art\_Project.jpg}}). The number of pixels of this image are $4001 \times 1757$, which is the largest available in the above URL.}  
\end{figure*}

\section{\label{s:pre}Pre-process of the image data}

\begin{figure}
\includegraphics[width=0.9\columnwidth]{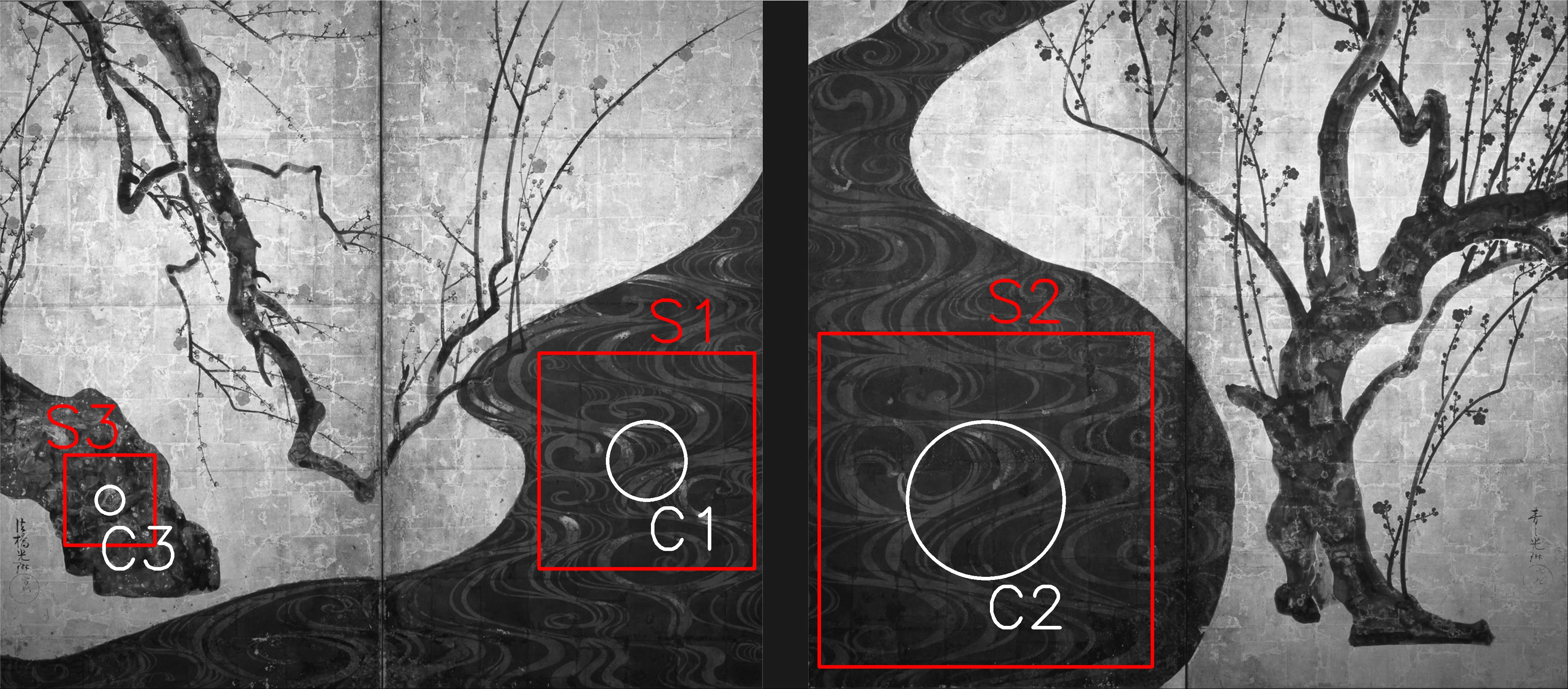} 
\caption{\label{f:mono}Gray-scaled image of Fig.~\ref{f:pic}. 
More specifically, the luminance $\theta(x, y)$ defined in Eq.~(\ref{def-theta}). 
The annotated circular regions, C1, C2, and C3, are used 
in the spectral analysis in Sec.\ref{s:spc}. 
The radii of C1, C2, and C3 shown here are set to $\sigma$'s listed in Table \ref{tg}.
The square regions, S1, S2, and S3, are used 
in the structure-function analysis in Sec.\ref{s:stf}.} 
\end{figure}

We first convert the color JPG image of \textit{Red and White Plum Blossoms} shown in Fig.~\ref{f:pic}
to a gray-scale image. We use the conversion formula,
\begin{equation}
 Y = 0.2125 R + 0.7154 G + 0.0721 B,
 \label{Y}
\end{equation}
from the red ($R$), blue ($B$) and green ($G$) channels of the image to the gray scale ($Y$).
Those $Y, R, G$ and $B$ are defined on each pixel at $(x, y)$.
The channel values are obtained here by using 
\texttt{cv2.split} of the library OpenCV (version 4.6.0) for computer vision
with Python (version 3.12.8).
The weight coefficients in Eq.~(\ref{Y}) are the standard ones known as ITU Recommendation BT.709-6, or Rec.709, 
for high definition television\cite{itu}. The conversion method is the same as
Ref.~\onlinecite{ma24}, but our coefficient values are slightly different from 
theirs. We check that the equal weights do not change the main results of this paper.
The value $Y(x, y)$ on each pixel is then transformed to a $8$-bit integer value which 
lies between $0$ to $255$. Let us now define the luminance $\theta(x, y)$ by
\begin{align}
 \theta(x, y)  = \frac{\lfloor Y(x, y) \rfloor_8}{255}.
 \label{def-theta}
\end{align}
Here $\lfloor \cdot \rfloor_8$ means the transformation to a $8$-bit integer. 
The resultant luminance is shown in Fig.~\ref{f:mono}. 
We compare the field $\theta(x, y)$ with the passive-scalar quantity advected by 
the homogeneous and isotropic turbulence (HIT), as proposed in Ref.~\onlinecite{ma24} for
\textit{The Starry Night}.

We use mainly the pixel coordinates in the following spectral and structure function analysis.
A pixel of the image shown in Fig.~\ref{f:pic}, or equivalently Fig.~\ref{f:mono},
can be regarded as a square by the following estimates.
The height of a pixel can be estimated as $156$ [{\rm cm}] $/1756$ [pixel] $\simeq 0.08884$ [cm$/$pixel].
We identify that the left-panel image extends horizontally from $0$ to $1945$ 
pixel and the right-panel image from $2062$ to $4000$ pixel.
The width of a pixel can be estimated as $172.2${\rm cm} $/1945$ [pixel] $\simeq 0.08853$ [cm$/$pixel] for the left panel
and as $172.2${\rm cm} $/1938$ [pixel] $\simeq 0.08885$ [cm$/$pixel] for the right panel.

\section{\label{s:spc}Spectrum via windowed Fourier transform}

\begin{figure}
\includegraphics[width=0.9\columnwidth]{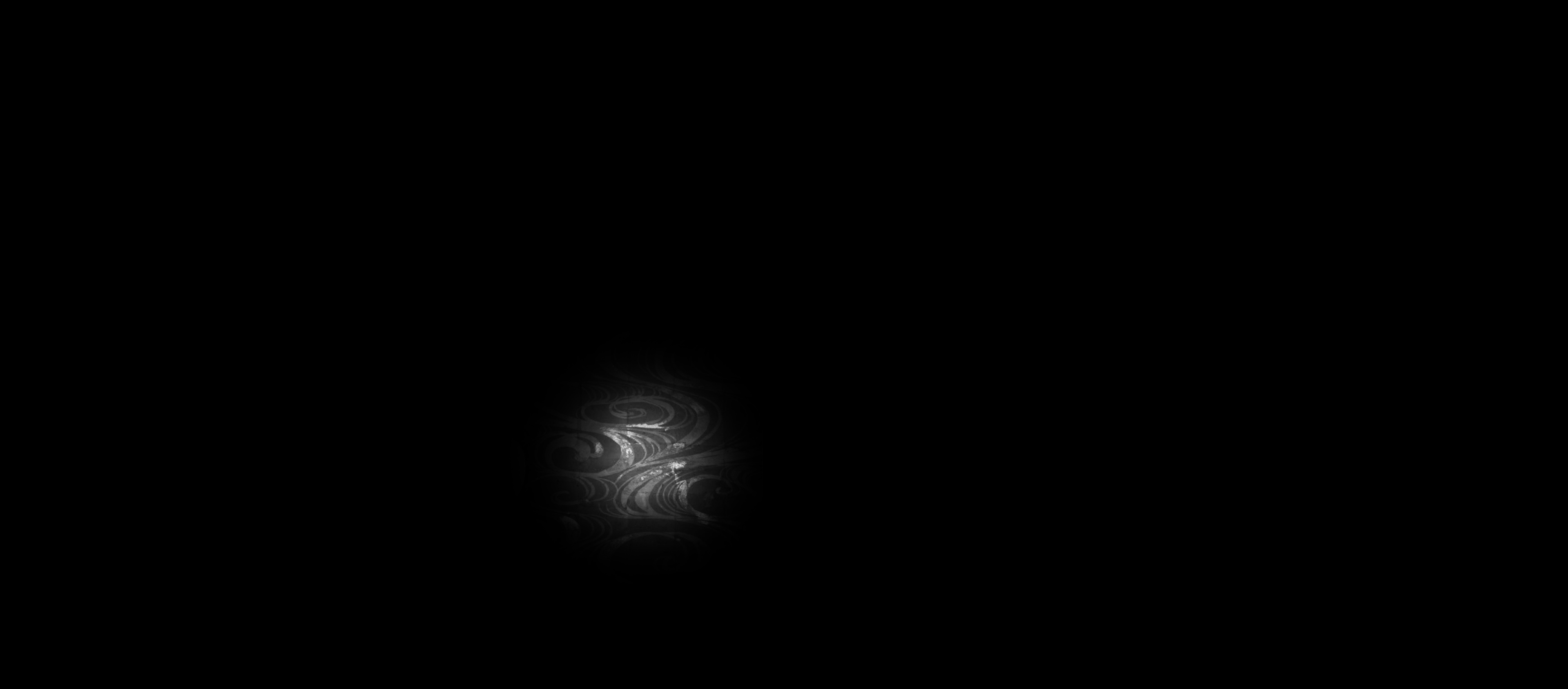}
 
\vspace*{0.5cm}
 
\includegraphics[width=0.9\columnwidth]{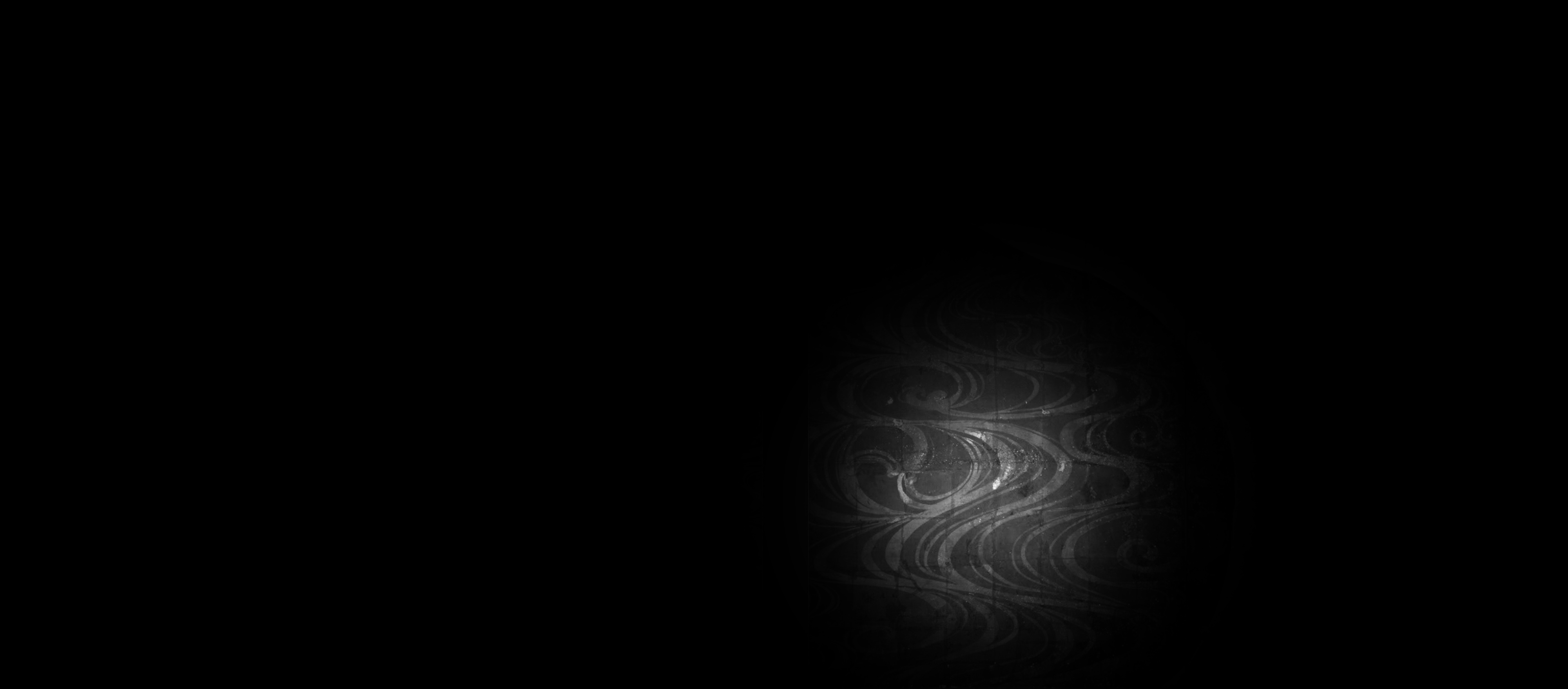}
 
\vspace*{0.5cm}
 
\includegraphics[width=0.9\columnwidth]{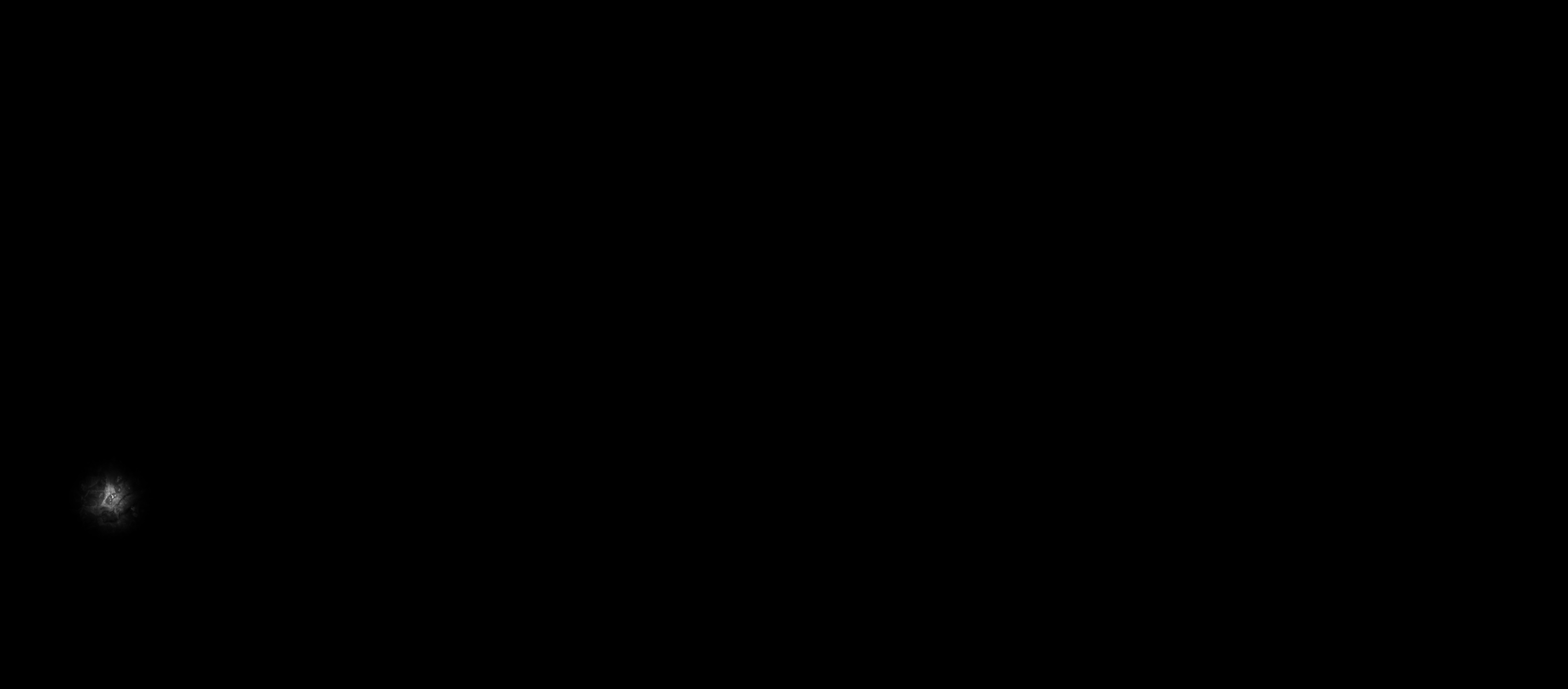} 
\caption{\label{f:win} Gaussian-windowed gray scale images for calculating the spectrum.
The top, middle and bottom panels correspond respectively to C1, C2 and C3 in Fig.~\ref{f:mono}.
The black color represents zero luminance after multiplied by the Gaussian function.
The non-zero luminance (gray or white) in this figure is enhanced than 
that of Fig.~\ref{f:mono} to highlight the effect of the windowing.}
\end{figure}

Let us calculate the spectrum of $\theta(x, y)$ in several parts
of the painting. We consider the parts including bulk of the river flowing 
between the white and red plum trees and the part other than the river, 
namely the bark of the white plum-tree trunk.
Since we focus on local parts of the painting,
we adopt the windowed Fourier transform, also known as the Gabor transform.
We consider three regions for calculating the spectrum. One of them is taken in the left panel
of the screen as depicted C1 in Fig.~\ref{f:mono} and the second one is taken in the right
panel as depicted C2 in the same figure.  The third one is taken in the left panel
as marked C3 in in the same figure. The regions C1 and C2 are in the river part
and C3 is on the trunk of the white plum.
The windowing is done with a two-dimensional Gaussian function,
\begin{align}
 \theta_w(x, y) 
 = \theta(x, y) 
 \frac{1}{2\pi \sigma^2} \exp\left[-\frac{|\vec{x} - \vec{x}_c|^2}{2\sigma^2}\right].
\end{align}
The parameter values for C1, C2 and C3 are listed in Table \ref{tg}.
The windowed luminance field $\theta_w(x, y)$ for the three regions are 
shown in Fig.~\ref{f:win}. Indeed they highlight the local regions.

The windowed field is then Fourier-transformed
and the spectrum is calculated in a standard manner,
\begin{align}
E_\theta(k) = \sum_{k \le |\vec{k}| < k + \Delta k} \frac{1}{2}|\hat{\theta}_w(\vec{k})|^2 \frac{1}{\Delta k}.
\end{align}
Here $\hat{\theta}_w(\vec{k})$ is the Fourier coefficient of the windowed luminance
and $\vec{k}$ is the wavevector. We take $\Delta k = 2\pi/(4 ~\text{[pixel]})$.

\begin{figure}
\includegraphics[width=0.9\columnwidth]{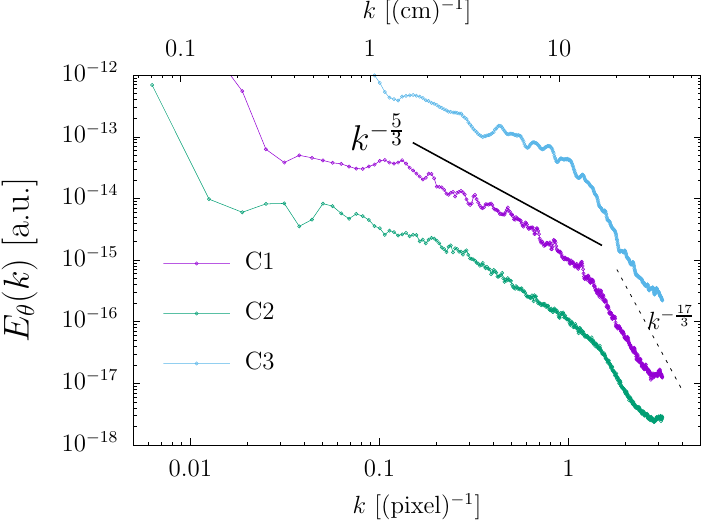}  
\caption{\label{f:spc} Spectra of the luminance 
calculated in the three circular regions, C1, C2, and C3, with the windowed Fourier transform. 
The solid line corresponds to the Obukuhov--Corrsin spectrum, $k^{-5/3}$,
in the inertial-convective range for the passive scalar advected by the homogeneous and isotropic turbulence.
The dashed line  corresponds to the Batchelor--Howells--Townsend spectrum, $k^{-17/3}$,
in the inertial-diffusive range for the passive scalar turbulence.
Here $1$ [pixel] corresponds to $0.089$ [cm].} 
\end{figure}

The resultant spectra for C1, C2 and C3 are plotted in Fig.~\ref{f:spc}. 
We observe the power-law behavior in the intermediate wavenumber range for 
all three cases.
This power law is consistent with the Obukhov--Corrsin (OC)
spectrum, $C \chi \epsilon^{-1/3}k^{-5/3}$, in the inertial-convective range 
for the passive scalar advected by HIT\cite{o49,co51,tl72}.
Here $C$, $\chi$, and $\epsilon$ are the non-dimensional Obukhov--Corrsin constant, 
the scalar dissipation rate, and the energy dissipation rate, respectively.

This $k^{-5/3}$ like power-law part of the luminance spectrum is followed by 
the steeply decreasing part in the larger wavenumbers. This is in contrast
to the spectrum found in \textit{The Starry Night} where the $k^{-5/3}$ part 
is followed by the shallower Batchelor spectrum $k^{-1}$ of the dissipative-convective range \cite{ma24}. 
The decreasing part for $k \ge 1$ [(pixel)$^{-1}$] in Fig.~\ref{f:spc} may correspond
to the dissipation-range behavior (non power-law behavior) of the passive scalar spectrum.
Or it may correspond to the Batchelor--Howells--Townsend spectrum, 
$\tilde{C} \chi \kappa^{-3} \epsilon^{2/3} k^{-17/3}$, of the passive scalar 
at low Schmidt numbers in the inertial-diffusive range \cite{bht59}, as indicated 
in Fig.~\ref{f:spc} by the dashed line.
Here $\tilde{C}$ and $\kappa$
are the non-dimensional constant and the diffusivity of the passive scalar, respectively.
The two possibilities suggest that the luminance is analogous to unit or low Schmidt number cases.
However, the range $k \ge 1 ~\text{[(pixel)]}^{-1}$ is too narrow to conclude
which one is more plausible. Furthermore, the smallest scales of the image can be
more affected by the image compression or other procedures of the image processing.
In this paper, we focus on the consistency with the OC spectrum.

In Fig.~\ref{f:spc} we do not show the luminance spectrum in the small wavenumber range. 
That part of the spectrum is the Gaussian function, namely $E_\theta(k) \propto \exp(-\sigma^2 k^2)$, due to the windowing.
The flat part of the luminance spectrum shown in Fig.~\ref{f:spc} ends roughly at $k = 2\pi/\sigma$ and the appearance 
of the power-law part in $k > 2\pi/\sigma$ indicates that the $k^{-5/3}$-like power law reflects the nature
of K\=orin's flowing water motif and pattern of the bark. 
Now we observed that the spectra of the painting shown in Fig.~\ref{f:spc} are close to $k^{-5/3}$.
However we cannot rule out that this consistency with the OC scaling law is just a coincidence,
since the pattern of the bark is not related to fluid motion. 
In the next section, we seek a different kind of consistency by calculating 
the even-order structure functions of the luminance up to 6th order.
\begin{table}
\caption{\label{tg}Parameters of the Gaussian window. The center coordinates $\vec{x}_c$ and the width $\sigma$ are given in terms of pixels of Fig.~\ref{f:pic}. Notice that the $y$ coordinate is increasing from top to bottom. The coordinate origin $\vec{x} = (0, 0)$ is the left-top corner of the figure. The coordinates of the right-bottom corner is $(4000, 1756)$.}
\begin{ruledtabular}
\begin{tabular}{lcc}
Region & $\vec{x}_c$ [pixels] &  $\sigma$ [pixels] \\
\hline
C1 & $(1650, 1175)$ & $100$\\
C2 & $(2515, 1275)$ & $200$\\
C3 & $(280, 1275) $  & $35$ \\
\end{tabular}
\end{ruledtabular}
\end{table}

\section{\label{s:stf}Structure function}
\begin{figure*}
\includegraphics[width=0.33\textwidth]{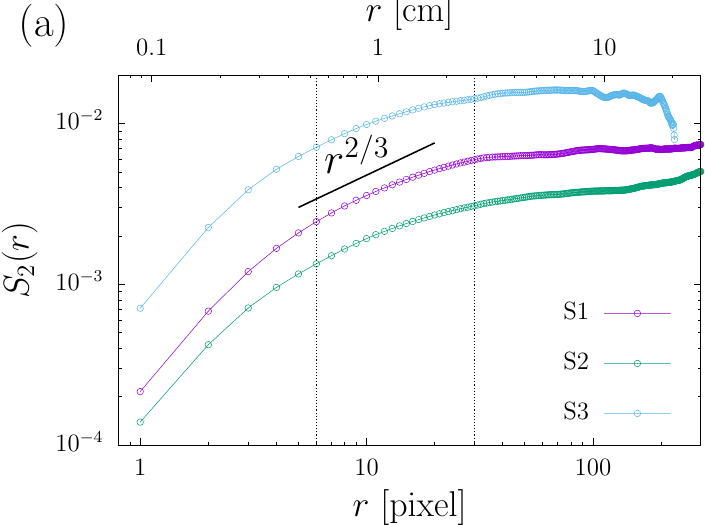}
\includegraphics[width=0.33\textwidth]{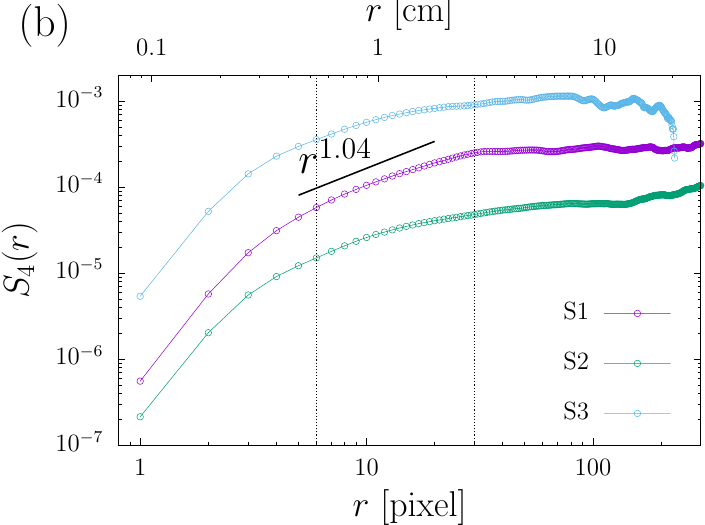}
\includegraphics[width=0.33\textwidth]{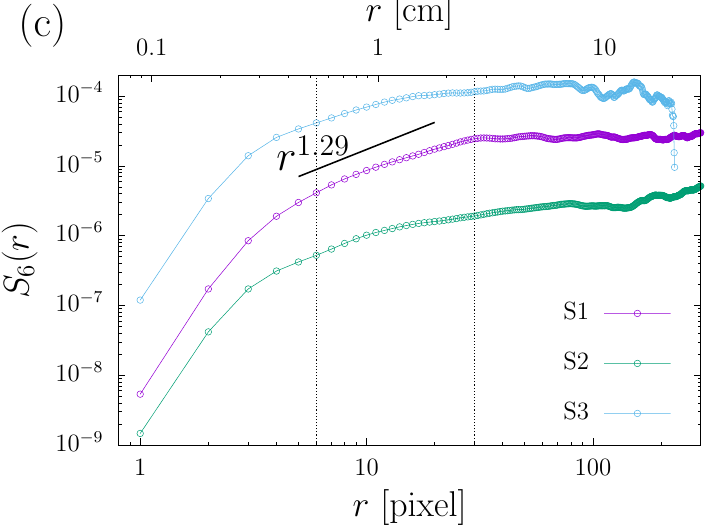} 
\caption{\label{f:stf}Even-order structure functions, $S_2(r)$ (a), $S_4(r)$ (b), and $S_6(r)$ (c),
 of the luminance calculated in the three square regions S1, S2 and S3. 
The solid lines denote the power-law exponents of the structure functions of 
the passive scalar. The exponent $2/3$ for the second-order one corresponds 
to the OC spectrum $k^{-5/3}$, for which the intermittency corrections are ignored.
The values of the other exponents ($1.04$ for $p=4$ and $1.29$ for $p=6$) 
are taken from the numerical study of the passive scalar in the homogeneous and isotropic
turbulence \cite{wg04}, so that the intermittency corrections are present.
As used in Fig.~\ref{f:spc}, one pixel corresponds to $0.089$ cm.}
\end{figure*}

First we define the luminance increment as
\begin{align}
\delta_r\theta = \theta(\vec{x} + \vec{r}) - \theta(\vec{x}).
\end{align} 
Then the $p$-th order structure function of the luminance is defined by
\begin{align}
 S_p(r) &= \langle (\delta_r \theta)^p \rangle,
 \label{stf} 
\end{align}
where $\langle \cdot \rangle$ denotes a spatial average. Here we assume
that the luminance is locally homogeneous and isotropic. Hence
there is only $r = |\vec{r}|$ on the left hand side of Eq.~(\ref{stf}).
We calculate the structure functions in the three square regions, S1, S2 and S3,
indicated in Fig.~\ref{f:mono}. The regions S1 and S2 are in the flowing river and the region S3
is in the bark of the plum-tree trunk.
The regions' coordinates are listed in Table \ref{ts}.
The spatial average is taken with respect to $\vec{x}$ in each region. The separation vector
is set to either $\vec{r} = (r, 0)$ or $(0, r)$. 
We take the averages over those two different directions of $\vec{r}$
and over all the instances in which both points $\vec{x}$ and $\vec{x} + \vec{r}$ 
are within each region.

The structure functions for $p = 2, 4$ and $6$ calculated in this manner 
are shown in Fig.~\ref{f:stf}. We here focus on those even-order functions since they 
converge more quickly than odd-order ones (the third-order function with and without the absolute value 
is discussed in the Appendix \ref{s:3}).
For the range of $r$ where we expect a power-law scaling, let us go back 
to Fig.~\ref{f:spc} in which the luminance spectrum is close to 
the OC scaling roughly in $0.2  \le k \le 1 $ (with the unit [$(\text{pixel})^{-1}$]). 
This wavenumber range corresponds to $6  \le r \le 30$ (with the unit [pixel]), which
are between the two vertical dotted lines in Fig.~\ref{f:stf}.

In this range of $r$, the second-order structure function, $S_2(r)$, in Fig.~\ref{f:stf}(a) 
is roughly consistent with the OC scaling, which is now $r^{2/3}$.
Let us recall that, if the OC scaling holds for the $p$-th order structure functions, 
the dimensional analysis yields $S_p(r) \propto r^{p/3}$ (see, e.g., Ref.~\onlinecite{gy13}).
Therefore $S_2(r)$ confirms that the luminance follows roughly the OC scaling.

\begin{figure}
\includegraphics[width=0.8\columnwidth]{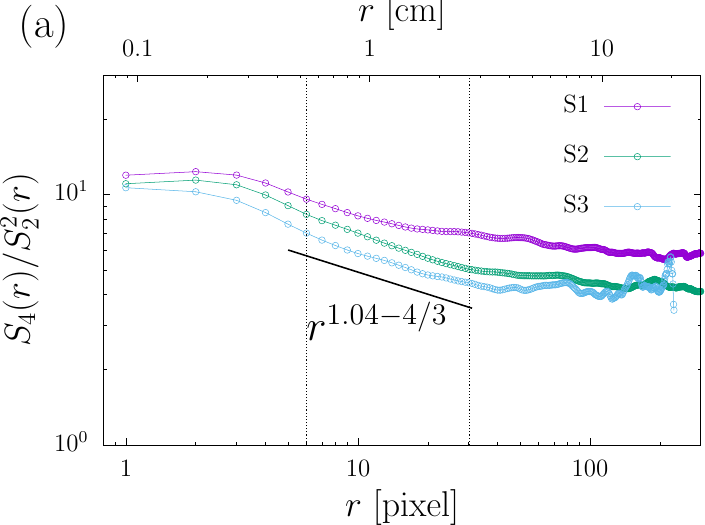} 

\vspace*{0.3cm} 
 
\includegraphics[width=0.8\columnwidth]{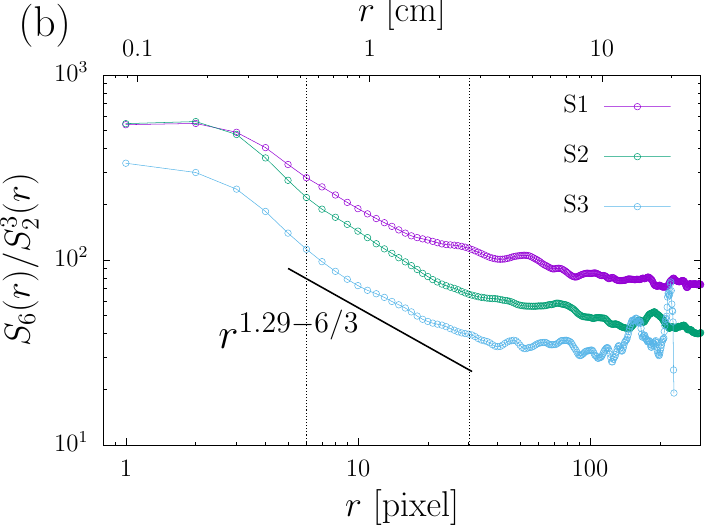} 
\caption{\label{f:flatness}Flatness $S_4(r)/S_2^2(r)$ (a) and hyperflatness $S_6(r)/S_2^3(r)$ (b)
of the luminance increment as a function of $r$. 
The solid lines represent the power laws obtained from the scaling exponents indicated in Fig.~\ref{f:stf}.}
\end{figure}

However, as shown in Fig.~\ref{f:stf}(b) and (c), the luminance structure functions for $p=2$ and $4$
differ from  $r^{4/3}$ and $r^{2}$, respectively. 
Instead they are closer to $r^{1.04}$ and $r^{1.29}$.
Those values of the exponents \cite{wg04} are so-called intermittent scaling exponents 
observed for the structure functions of the passive scalar advected in HIT. 
By intermittent\cite{f} we mean
that the exponent of $S_p(r)$ is different from $p/3$ that is obtained by the dimensional 
argument\cite{o49,co51,tl72} leading to the OC spectrum $k^{-5/3}$.
It is known that, for the passive scalar turbulence, the power-law exponent for $p=2$ 
differs from $2/3$ only slightly. However, for $p > 2$, the difference becomes larger, a phenomenon
known as the intermittency\cite{f,rmp,w00,gy13}.
Those differences, called intermittency corrections, 
are highlighted in the flatness $S_4(r)/S_2^2(r)$ and the hyperflatness $S_6(r)/S_2^3(r)$
of the luminance increment, as shown in Fig.~\ref{f:flatness}. 
Variations are present depending on the regions, S1, S2 and S3.
But, overall, their $r$-dependence in the scaling range indicates that 
the intermittency of the luminance is approximately the same as the passive scalar advected by HIT.

Aiming at observing a cleaner power-law scaling, we use the SO(2) decomposition to extract the 
isotropic sector of $S_p(r)$\cite{bp05}. However the
isotropic sectors (not shown here) are basically the same as those shown in Fig.~\ref{f:stf}, indicating
that the anisotropic effects are not strong. Therefore we consider that contamination
making the graphs of $S_p(r)$ in Fig.~\ref{f:stf} not perfectly straight is primarily 
due to the inhomogeneity.

\begin{figure}
\includegraphics[width=0.8\columnwidth]{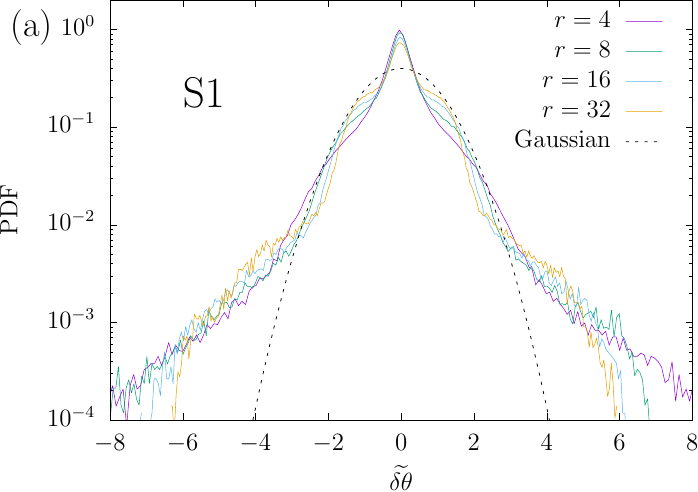}

\vspace*{0.3cm}  
 
\includegraphics[width=0.8\columnwidth]{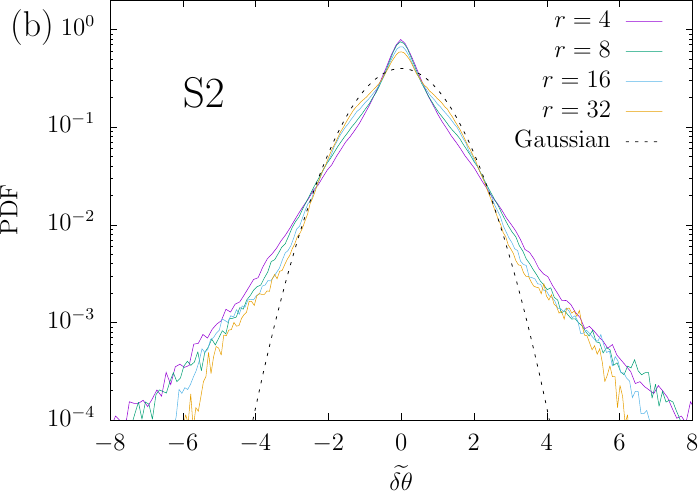}

\vspace*{0.3cm}  
 
\includegraphics[width=0.8\columnwidth]{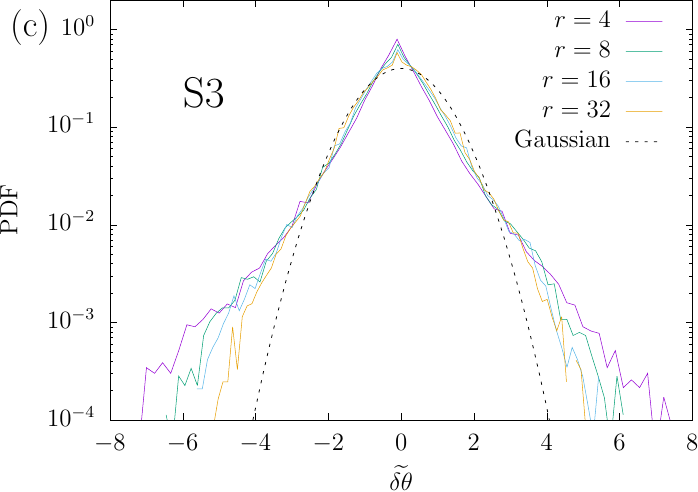} 
\caption{\label{f:pdf} Probability distribution functions of the luminance increments 
in the region S1 (a), S2 (b), and S3 (c) for several separation $r$'s in the scaling 
range of the structure functions. The values of $r$'s are given in pixels.
The increment denoted by $\widetilde{\delta \theta}$ here is the normalized 
increment such that its mean and standard deviation are zero and unity, respectively.} 
\end{figure}

The intermittency of the structure functions can be a manifestation of 
the multifractality\cite{f} of the luminance. It implies that the scale invariance of the luminance
is broken. This can be observed in the probability  distribution function (PDF) 
of the luminance increment for different $r$'s. We show the PDFs
in the S1, S2 and S3 regions in Fig.~\ref{f:pdf} for the normalized increment 
$\widetilde{\delta_r \theta} = [\delta_r \theta - \langle \delta_r \theta \rangle]/[\langle (\delta_r \theta)^2 \rangle - \langle \delta_r \theta \rangle^2]^{1/2}$.
Obviously the PDFs are not Gaussian, having fat tails. More importantly,
the normalized PDFs for several $r$'s in the scaling range are not collapsed 
to one master curve, although the differences between the PDFs are small. 
Thus the scale invariance is broken. This non-collapsed behavior of the non-Gaussian PDF 
is similar to that of the passive scalar in HIT\cite{w00,wg04,gy13}. We notice that the tail of the PDF 
for S3 (the bark) in Fig.~\ref{f:pdf}(c) is less pronounced than those for S1 and S2 (the swirls). 
This tail difference does not result in sizable differences in the structure functions 
up to the 6th order shown in Fig.~\ref{f:stf}.

In summary, we find that K\=orin's flowing water motif and the bark of the plum-tree trunk 
respect approximately the physical law of the passive scalar advected by HIT. By the law, we mean not only the OC scaling of the second 
order quantities, but also the intermittent exponents of the 4th- and 6th-order structure functions.
Those intermittent exponents of the passive scalar turbulence are known to be universal in the sense that 
they do not depend too much on details of the system, such as how to inject the scalar quantity into the turbulent 
flow \cite{rmp,gy13}. 
Such agreement for this highly stylized depiction of flow (S1 and S2) is likely not trivial at all 
and thus comes as a delightful surprise to us. 
The agreement for the bark (S3) appears puzzling since
it is not a depiction of a flow. In the next section we will argue as one possibility 
that it has something to do with the painting technique used for the bark. 

\begin{table}
\caption{\label{ts}Coordinates of the square regions, S1, S2 and S3, for the structure functions. 
The centers are the same as those of C1, C2 and C3 used in the windowed Fourier transform.}
\begin{ruledtabular}
\begin{tabular}{lcc}
Region & Center [pixels] &  Side length [pixels] \\
\hline
S1 & $(1650, 1175)$ & $550$\\
S2 & $(2515, 1275)$ & $850$\\
S3 & $(280, 1275)$ & $250$ 
\end{tabular}
\end{ruledtabular}
\end{table}

\section{\label{s:concl}Concluding discussion}

We have analyzed the luminance of the flowing river and the bark of the white plum-tree trunk 
in \textit{Red and White Plum Blossoms}.
The spectrum of the luminance with the windowed Fourier transform followed closely $k^{-5/3}$ in
the two different regions of the river and the bark region. 
This suggests that the luminance spectrum is consistent with 
the Obukhov--Corrsin spectrum of the passive scalar advected by HIT.
To examine this consistency further,
we calculate the even-order structure functions of the luminance up to the 6th order 
and found that they showed the power-law behavior although its range
was limited. The power-law exponents of the luminance structure functions were also approximately 
consistent with the exponents with the intermittent exponents of the passive scalar in turbulence. 
Therefore we conclude that the flowing-water pattern and the depiction of the bark of the white 
plum-tree trunk of \textit{Red and White Plum Blossoms} follow 
the universal (intermittent) scaling law of the passive scalar in the turbulent flow.

\subsection{Robustness of the scaling}
Is the quantitative agreement between the stylized artwork and the natural phenomenon real or not? 
Of course, it can be an artifact in the process of making an electronic image such as 
the image processing, the photographic lighting and so forth, which are not related to 
the painting itself.
To address this issue, we take two other JPG images
of \textit{Red and White Plum Blossoms}, which are supposedly independent from the one 
shown in Fig.~\ref{f:pic}. Here we limit ourselves to the river regions where the number of
data points are relatively large. We analyze them in exactly the same manner. 

One of the images is the one on the web site of the MOA museum of art\cite{MOAjpg}
and the other one is on the Japanese Wikipedia on \textit{Red and White Plum Blossoms}\cite{JAWjpg}.
We call the former MOA image and the latter JAW image. We then call the image shown in Fig.~\ref{f:pic}
GAC image.  The number of pixels of the MOA and JAW images are smaller
than the GAC image ($1500\times658$ for the MOA image and 
$2077\times 925$ for the JAW image). Consequently, the scaling laws in the MOA and JAW images are
less clear than what we observed in the GAC image. 
For the spectra in the regions equivalent to C1 and C2, the MOA and JAW images are
consistent with the GAC image (here we take approximately the same regions for each image). 
For the structure functions in the region S1, the MOA image is not consistent
with the GAC image, while the JAW image is consistent. In the region S2, only the 2nd-order structure functions
of the MOA and JAW images are consistent with that of the GAC image. The 4th- and 6th-order
structure functions in S2 of the MOA and JAW images are not consistent with that of the GAC image.  
The inconsistent structure functions are nearly $r$-independent except for the smallest few pixels, 
suggesting that the luminance have sharp spikes which are much higher 
than the background in the MOA and JAW images.  Indeed, in the regions we observe
the inconsistency, such spikes are found in the MOA and JAW images. The GAC image does not have 
comparable spikes.

This comparative little study indicates that the scaling property of the luminance
is indeed affected by the image processing. However, if the luminance is sufficiently less spiky, or smooth enough,
the scaling property seems to be robust. Of course, the number of pixels (resolution) is important for the smoothness, 
as the spectra of the MOA and JAW images in the region C2 do not have the hypothetical dissipation range, namely 
the part decreases faster than $k^{-5/3}$. Our conclusion here is that the less-spiky and high-resolution
GAC image better reflects the nature of \textit{Red and White Plum Blossoms} and that the turbulent scaling law 
we observed in the GAC image is likely robust.

\subsection{Why consistent?}
The last question would be why and how K\=orin created the flowing water pattern, or \textit{suiryu moy\=o} in Japanese,
and the bark of the plum-tree trunk, resulting in approximately the same intermittent scaling law of the passive scalar turbulence.

Now let us go back to \textit{The Starry Night} that has the similar quantitative consistency with the real
passive scalar turbulence\cite{ma24}. In Ref.~\onlinecite{ma24}, it was argued that van Gogh's
very careful observation as a master of the post-impressionism led to the reproduction of the natural phenomenon.
Furthermore the Batchelor scaling found in \textit{The Starry Night} at the small scales was ascribed
to the painting materials having a high Schmidt number\cite{ma24}.
For \textit{The Great Wave off Kanagawa}, it was also concluded that Hokusai's talent as an observer enabled
him to capture and reproduce the physics of the waves \cite{dudley13}.

In the case of K\=orin, the same argument about his ability of observation applies without doubt.
We first consider the bark of the trunk of the white plum tree (the region C3 and S3). 
The painting technique used for this part is well-known as \textit{tarashikomi} in Japanese 
\cite{lippit07,moma,chuo}, which is developed by the K\=orin school. 
In this technique, the overlaied paint is placed while the underlaied paint is still wet. 
In the bark part of \textit{Red and White Plum Blossoms}, its texture and lichens on it 
are depicted with \textit{tarashikomi}\cite{chuo}.
Although the bark does not look like a passive scalar vigorously stirred by a turbulent flow,
the \textit{tarashikomi} technique can give rise to turbulence locally in the 
paints to some extent. As a result, we observed the turbulent scaling law there. 
This is an explanation we propose for the bark region. 
It is in line with the argument for the Bachelor scaling in \textit{The Starry Night}\cite{ma24}.
Still, how the turbulent stirring we propose, if it is true, reproduces the turbulent scaling law when the paints are dried
remains to be understood. 
Now, for a possible future study, let us estimate the dissipation scale.
From the luminance spectrum shown in Fig.~\ref{f:spc}, the end of the $k^{-5/3}$-like part is 
estimated as $k \simeq 15$ [(cm)$^{-1}$]. It corresponds to $0.42$ [cm], which can be regarded 
as the Kolmogorov dissipation length for the case of unit Schmidt number or the Obukhov--Corrsin length for 
the case of the low Schmidt number.

The same explanation may not be applicable to the flowing water pattern in the regions C1 and C2 
or S1 and S2 since the painting technique used there is not known and a matter of debate.
A recent study\cite{kajima} involving present-day artists
using similar materials argues that the flowing pattern
is likely drawn freehand in one sitting and that use of a stencil could not
explain the expressed sense of momentum (but use of a stencil is not completely ruled out).
Paper marbling (\textit{suminagashi} in Japanese) or water transfer printing is ruled out.
Indeed, looking at Fig.~\ref{f:pic} or enlarged photographs of various parts of the painting in Ref.~\onlinecite{chuo},
we can imagine that K\=orin drew without hesitation the lines of the flowing patterns.
One possibility is that finishing in a single sitting with some momentum 
can induce an effect similar to \textit{tarashikomi}, 
namely, local turbulent stirring, which leads to the scaling law we observed.
Here we infer that the effect results in variations in ink density within the brown curves of 
the swirling pattern. 

Nevertheless, for the flowing water pattern in particular, there is another possibility.
That is, the turbulent scaling law there is not due to the turbulent stirring of the painting material,
but due to K\=orin's depiction. 
Specifically, his irregular hierarchy of the flow pattern ultimately ended up reproducing it.
This is a far fetched speculation, though.
Further investigation of K\=orin's construction of the flow with much higher resolutions
will tell us whether it is likely.
If it is, then his way of depiction may leads a method that is in the same spirit of the recent mathematical construction of turbulent 
weak solutions to the Euler equations, known as the convex integration \cite{bdsv}.
If it succeeds, then we see an example of the provocation made in Ref.~\onlinecite{w22}: ``Artists can complement scientific analysis and help in developing a comprehensive theory of turbulence''.

As a final remark, there is still a possibility that 
the scaling law of the parts of \textit{Red and White Plum Blossoms} has nothing to do with
fluid turbulence. If this is the case, it could imply that there is a universality class of scalar fields
having the same intermittent scaling law and that the passive scalar turbulence and the parts of 
\textit{Red and White Plum Blossoms} belong to this universality class.

\begin{acknowledgments} 
I thank the MOA Museum of Art, 
the Wikimedia Commons and the Google Art \& Culture initiative
for making the artwork available in the high-resolution JPG format.
I gratefully acknowledge Kohei Okuyama for discussions, 
one referee for critical comments
and the other referee for suggesting analysis of the bark on the trunk 
of the left plum tree (the regions C3 and S3). 
This work was supported by JSPS KAKENHI Grant Number JP19K03669
and Number JP23K03247.
\end{acknowledgments}

\section*{Data Availability Statement}

The JPG image of \textit{Red and white plum blossoms}, which we 
analyzed here, is available at Wikimedia Commons, 
\url{https://commons.wikimedia.org/wiki/File:Ogata_Korin_-_RED_AND_WHITE_PLUM_BLOSSOMS_(National_Treasure)_-_Google_Art_Project.jpg}.
The file was uploaded to Wikimedia Commons on November 26th, 2013 as written in the above web page.
We retrieved the file on October 22nd, 2024 for the present analysis.

\appendix

\section{\label{s:3}Third-order structure function}
\begin{figure}
\includegraphics[width=0.8\columnwidth]{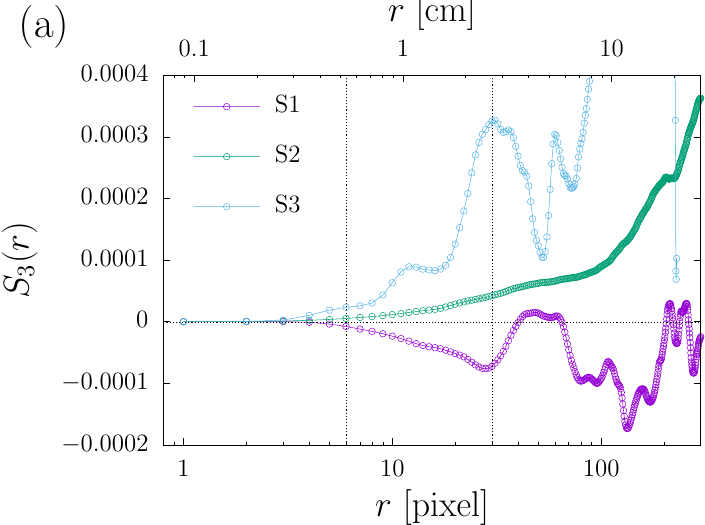}

\vspace*{0.3cm} 
 
\includegraphics[width=0.8\columnwidth]{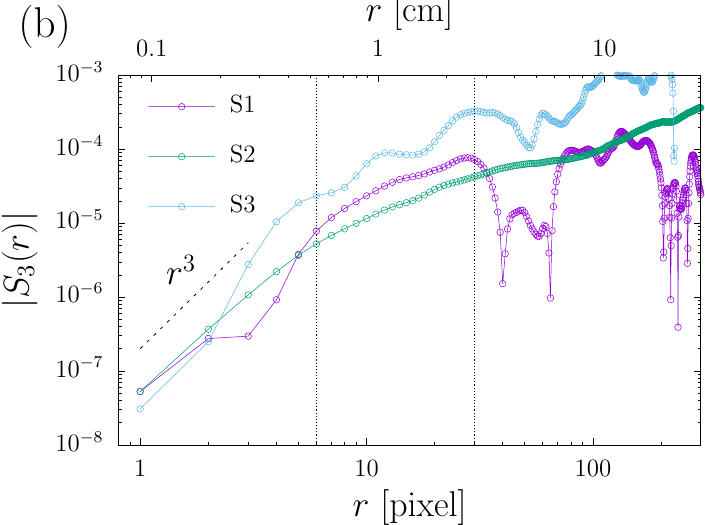}

\vspace*{0.3cm}  

\includegraphics[width=0.8\columnwidth]{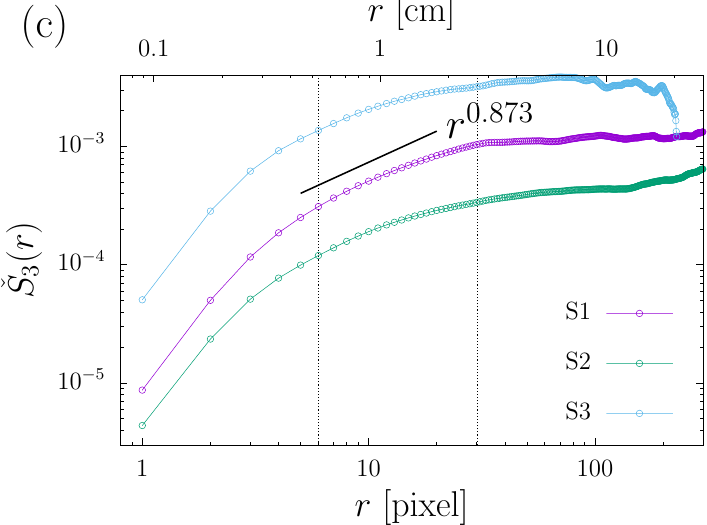} 
 
\caption{\label{f:third}Third-order structure function $S_3(r)$ in the linear-log coordinates (a) and its absolute value in the log-log coordinates (b). Third-order structure function of the absolute value of the luminance increment $\check{S}_3(r) = \langle |\delta_r \theta|^3 \rangle$ is also shown in (c). The exponent $0.873$ in the panel (c)  is taken from Ref.\onlinecite{wg04} for the third-order moments of the absolute value of the passive scalar increment.}
\end{figure}

In this appendix we consider the 3rd-order structure function of the luminance
in the square regions, S1, S2 and S3. 
As we mentioned in Sec.~\ref{s:stf}, 
convergence of the odd-order moments involving cancellation 
requires more data points than the even-order ones do.
Accordingly, the results presented here may not be sufficiently 
converged. Nevertheless, we plot $S_3(r)$ in Fig.~\ref{f:third}(a). 
The figure shows that its sign varies: it is positive in S2 and S3, but it is not
definite in S1. The wiggly behaviors of the curves for S1 and S2 indicate
the lack of convergence, though.
In the numerical simulation of the passive scalar turbulence, 
the skewness of the passive scalar increment is found to be 
a small positive constant over the inertial-convective range \cite{wg04}.
However, since the forcing added to the passive scalar equation in Ref.\onlinecite{wg04} 
is a large-scale random forcing, it is argued that the skewness should vanish, 
provided that the turbulence in the large scales is isotropic enough\cite{wg04}. 
If the passive scalar is maintained by the mean scalar gradient, it is known that 
the skewness is not zero \cite{w00}. 
We do not pursue our comparison in his direction however.

The most important 3rd-order moment in the passive scalar turbulence
is the Yaglom $4/3$ law that involves the velocity increment.
For the painting, if there is some object that can be interpreted as the velocity field, 
then we can study whether the Yaglom $4/3$ law\cite{yaglom, w00, gy13}, 
$\langle \delta_r u (\delta_r \theta)^2 \rangle = - (4/3) \chi r$,
holds in the scaling range. Here $\delta_r u$ is the increment of the velocity-like object.
The factor $-4/3$ is for the three spatial dimensions.
Testing against the Yaglom law is a reliable examination of the consistency,
for it is an exact relationship for the passive scalar turbulence in the inertial-convective range.
Unfortunately we do not have such a velocity-like object.

As shown in Fig.~\ref{f:third}(b), the log of $|S_3(r)|$ suggests power-law behavior.
Then we calculate the the third-order moments of the absolute value of the luminance, 
$\check{S}_3(r) = \langle |\delta_r \theta|^3 \rangle$, which is shown in 
Fig.~\ref{f:third}(c). Thanks to the absolute value, the curves become much less wiggly and 
the scaling exponents of $\check{S}_3(r)$ is close to the corresponding exponents, 0.873,
of the passive scalar turbulence \cite{wg04}, adding another piece of consistency of the painting
to the turbulence. 

%\noindent
\section*{\textit{Note added in proof}}
A referee kindly pointed out the reference in vision science,
D.J. Tolhurst, Y. Tadmor, and T. Chao, "Amplitude spectra of natural images", Ophthalmic and Physiological Optics \textbf{12}, 229--232 (1992).
In this reference, the luminance spectra of 135 photographs of diverse objects 
such as standing two persons, horses, natural scenery, a house and so forth, were analyzed. It is shown that
their luminance spectra exhibit power-law scaling with the exponents
ranging from $-2$ to $-0.6$, which include $-5/3$.
This result demonstrates that agreement of the the power-law exponents of the spectra
between a given image and the scalar-variance spectrum of turbulence does not necessarily imply
a fundamental connection of the image to the passive scalar turbulence.
In the present paper, we showed the agreement of
not only the scaling of the luminance spectrum, but also the scaling of the structure function.
Whether the structure functions of natural images exhibit a similar intermittent scaling law 
of the turbulent flow is an interesting subject to study.

More specifically, Tolhurst \textit{et al.} analyzed the amplitude spectrum of the luminance,
$A(k) = \int_0^{2\pi} |\hat{\theta}(\vec{k})| k d\varphi /[\int_0^{2\pi} k d\varphi]$ in the two-dimensional wavenumber space.
Here $\varphi$ is the polar angle in the wavenumber space.
The amplitude spectrum is related to the luminance spectrum $E_\theta(k) =  \int_0^{2\pi} (1/2) |\hat{\theta}(\vec{k})|^2 k d\theta$ 
as $E_\theta(k) = \pi k A(k)^2$ for the isotropic case. 
Tolhurst \textit{et al.} found $A(k) \propto k^{-\alpha}$ with $0.8 \le \alpha \le 1.5$. This corresponds to $E_\theta(k) \propto k^{-\beta}$
with $0.6 \le  \beta = 2\alpha - 1 \le 2$ by assuming the isotropy of the Fourier modes $\hat{\theta}(\vec{k})$.
(This note is not included in the published version).

\nocite{*}
%\bibliography{aipsamp}% Produces the bibliography via BibTeX.
\bibliography{korin}% Produces the bibliography via BibTeX.

\end{document}